# Bright quasi-phasematched soft x-ray harmonic radiation from Argon ions

M. Zepf[1*], B. Dromey[1], M. Landreman[2], P. Foster[3], S. M. Hooker[2]

[1] *Department of Physics and Astronomy, Queens University Belfast, BT7 1NN*
[2] *Clarendon Laboratory, Oxford University, OX1 3PU, UK*
[3] *Central Laser Facility, CCLRC Rutherford Appleton Laboratory. Chilton, Didcot,OX11 0QX, UK*
*[*]m.zepf@qub.ac.uk*

Selective enhancement ($>10^3$) of harmonics extending to the water window (~4nm) generated in an argon gas filled straight bore capillary waveguide is demonstrated. This enhancement is in good agreement with modeling which indicates that multimode quasi-phasematching (*MMQPM)* is achieved by rapid axial intensity modulations caused by beating between the fundamental and higher order capillary modes. Substantial pulse-energies (>10nJ per pulse per harmonic order) at wavelengths beyond the carbon K-edge (~4.37nm, ~284eV) up to ~360eV are observed from argon ions for the first time.

High Harmonic Generation (HHG) [1,2] using ultra-fast laser sources is an attractive route to generating coherent beams of extreme ultra-violet (XUV) radiation in the 20-400eV spectral region [3, 4, 5]. The key advantages of the HHG approach are the high beam quality, ultra-short pulses ranging from femtoseconds (fs) to attoseconds [6] and the comparative ease with which compact, commercially available femtosecond lasers can be converted to the XUV.

The most significant barrier to widespread application of HHG is the low conversion efficiency, and consequently low output that can be achieved. While reasonable conversion efficiencies in the range of $E_{harmonic}/E_{Laser}= 10^{-5} – 10^{-6}$ ($E_{harmonic}$ and $E_{Laser}$ are the energies in a single harmonic and laser pulse respectively) have been achieved at photon energies <40eV [3], the reported conversion efficiency at wavelengths >200eV has been much lower ($10^{-11}$) [4]. This is due to the fact that the highest efficiencies can only be achieved by phasematching the harmonic production process throughout the length of the generating medium, thereby ensuring that the harmonic signals from different parts of the medium interfere constructively. Under these circumstances, the intensity of the $q^{th}$ order harmonic will grow as $I_q \sim L_m^2$ and $I_q \sim N^2$ under phasematched conditions (where $L_m$ is the medium length and $N$ is the number density of emitters). Phasematching requires that the propagation vectors satisfy

$$\Delta k = k_q - q k_0 \qquad (1)$$

where $\Delta k$ is the wavevector mismatch, $k_q$ and $k_0$ the $q^{th}$ harmonic and laser wavevector respectively. In practice this condition cannot be met at high photon energies, because generating high order harmonics [2] requires that very high laser intensities are used. This results in a large ionisation fraction (>0.5) at the peak of the pulse, which leads to a large negative dispersion that cannot be compensated by the positive dispersion of neutral gas [3].

Here we report on a recent experiment performed on the Astra laser at the Rutherford Appleton Laboratories [7] that has demonstrated substantial enhancement (>$10^3$) of the HHG signal – particularly in the water-window (~4nm). This enhancement is interpreted as being the first observation of a novel scheme for quasi-phasematching (QPM) of the harmonic signal generated in gas filled capillary wavguides, multimode quasi phase-matching (*MMQPM*), in which QPM arises from the strongly modulated intensity profile formed when multiple waveguide modes are excited [9].

QPM [8, 10, 11] provides a promising alternative to current phasematching schemes, by eliminating the need to achieve $\Delta k=0$. Instead, QPM relies on periodically suppressing the HHG process in regions, which would contribute destructively to the harmonic signal, i.e. the HHG source term is modulated with a QPM period

$$L_{QPM} = \frac{2\pi m}{\Delta k} \quad (3)$$

where $m$ is the order of the process. As a result the signal can grow substantially above that generated from a single coherence length, albeit more slowly than with true phase-matching. In principle, QPM can be achieved in a variety of ways, e.g. by varying the medium density or by modulating the intensity of the fundamental driving laser. The latter approach is highly promising, since the strong intensity dependence of HHG implies that a relatively small reduction in the intensity of the fundamental results in a substantial reduction of the harmonic signal. Substantial enhancements are possible with periodic variations as low as a few percent [8, 10].

One platform for achieving a suitable intensity profile for QPM is via intensity modulations present in a hollow-core capillary with multiple excited modes [9, 12]. In the idealised case of only two excited modes, it is easy to see that the resulting on-axis intensity profile in a capillary will display periodic intensity modulations: the two modes, the fundamental mode, $j=1$, and the $j^{th}$ mode propagate with different $k$, vectors $k_1$ and $k_j$, respectively and consequently produce a regular intensity beat pattern of period

$$L_{QPM} = \frac{2\pi}{k_1 - k_j} \quad (4)$$

QPM will then occur when $\Delta k = k_1 - k_j = 2\pi/L_{QPM}$, from Eq. (3), which can be achieved by choosing appropriate values for the peak intensity and the gas density.

In practice it is difficult to achieve excitation of only two waveguide modes owing to the imperfect mode matching at the capillary input plane and mode coupling within the capillary due to ionisation [12]. One might assume that the intensity modulation resulting from multiple modes beating in capillary would be detrimental to achieving QPM. However, close inspection of the resulting on-axis intensity profile shows that the spacing of the peaks in the axial intensity are determined by the highest-order, significantly contributing mode in the capillary [9]. In simple terms, the main difference between two modes and multiple mode beating is that some peaks are 'missing' and hence the overall signal growth is reduced – but still substantial enough to allow significant gains to be made over fairly short medium lengths.

This significantly modulated intensity in the capillary has a major benefit for the quasi-phasematching scheme – while instantaneously high at points of high intensity, the average ionisation, $Z^*$, is significantly reduced resulting in a reduced $\Delta k$, which in turn requires a longer - and crucially, easier to achieve - $L_{QPM}$ to compensate (typically for soft x-ray harmonics $L_{QPM} < 50 \mu m$ in a high ionisation limit).

This experiment was performed on the Astra laser at the Rutherford Appleton Laboratory. Astra delivered ~40fs pulses limited to a maximum of 50mJ on target for this experiment. An f=1m focusing lens was used to couple the laser into the capillary entrance (bore radius, $a$, ~90$\mu$m, up to 15mm long). The capillary coupling parameter was varied by aperturing the beam from >20mm to 10mm diameter. The capillary was filled with argon gas via two laser-machined entrance holes resulting in a constant pressure region of 10mm length. The harmonic radiation was detected using an ANDOR

XUV CCD detector coupled to a flatfield grating spectrometer with an angular acceptance of ~5mrad × 5mrad.

For a Gaussian beam optimal coupling into the fundamental mode of the capillary is achieved for the vacuum coupling parameter, $\chi$, given by [12, 13]

$$\chi = \frac{w_0}{a} = 0.64 \tag{5}$$

where $w_0$ is the $1/e^2$ radius of the incident intensity profile and $a$ is the capillary bore radius. For $\chi \neq 0.64$, or a non-Gaussian pulse profile, more and more incident energy is excited into higher order modes. Under typical experimental conditions the transverse profile of the incident beam will more closely resemble an Airy profile, due to the fact that for efficient energy extraction in high power laser systems the near-field transverse intensity profile is approximately top-hat pulse.

A lineout of an experimentally measured laser focal spot is shown in Figure 1a) (solid line), corresponding approximately to an Airy profile with $\chi \approx 0.2$ for the central bright maximum (dotted line). The calculated on-axis intensity profile $I(z)$ for a 1cm long evacuated capillary (no gas present) with $a \sim 90 \mu m$ for coupling corresponding to the dotted profile in Figure 1a) is shown in Figure 1b). The rapid fluctuations in the on-axis intensity arise from beating between the fundamental and higher-order capillary modes. The peaks in the intensity profile are spaced by multiples of the beat period, $L_B$, allowing quasi phase-matching through the *MMQPM* process [9]. The expected growth of the harmonic signal for $q=201$ for this intensity profile, with pressure, $p \sim 20$mbar Ar and an incident energy of ~7mJ, leading to a peak intensity $I_{max} \sim 1 \times 10^{15} Wcm^{-2}$ in the capillary, is shown in Figure 1 c).

The signal growth for $q=201$ from an optimised two mode *MMQPM* scheme (i.e. pressure and intensity matched to satisfy Eq. 3 and 4 for modes $j=1$ and $j=20$ beating with equal intensity and ignoring mode attenuation, giving $L_B = L_{QPM} = 200 \mu m$ for $a \sim 90 \mu m$ [9]) is represented by the dashed lined in Figure 1b). The harmonic order $q=201$ is generated from Ar ions which, as result of higher ionisation potential, has a higher harmonic cut off than that of neutral Ar [5]. It should be noted that the rapid growth of HHG over the first few mm of the capillary for the intensity profile in Figure 1b) compared to that for the two mode *MMQPM* scheme (Eq. 4) is due to the constructive interference of multiple modes leading to an increased on-axis intensity above that achieved in the idealised two mode case.

Strong enhancement of the HHG signal in the water-window region was observed experimenatlly for a vacuum coupling parameter of $\chi \sim 0.2$ and $p \sim 20$mbar Ar. Figure 2 compares the background subtracted signals (detector limited) obtained for quasi-phasematched conditions (20mbar Ar, red and blue traces) and mismatched conditions (42mbar Ar, black trace), displaying an enhancement $>10^3$. The dip in harmonic intensity at 4.37nm is due to the transmission curve of the 0.1µm Al and 0.2µm CH filter used to block optical light. The maximum harmonic order detected was $\sim 231^{st}$ (>360eV, 3.43nm second order diffraction). Figure 2 also shows the short-term reproducibility of the data; the red and blue curves correspond to data taken within a 3-minute interval. No signal was observed in the water-window when the pressure was changed to mismatched conditions.

The lower limit of the conversion efficiency (as described in Figure 2) is estimated to be $>10^{-6}$ per harmonic at 300eV after correcting for spectrometer response and including only the central, bright signal. This implies a photon flux of $>10^{10}$ per harmonic peak per second (10Hz system) and a peak brightness of $>10^{21}$photons/s/mm$^2$/mrad$^2$ in 0.1% bandwidth, making this the brightest source of water window region x-rays from HHG observed to date.

One important point of note is that this is the first observation of QPM (albeit imperfect i.e. peaks missing) in ions with $m=1$. Under ordinary conditions, exploiting the higher ionisation potential of ion species for increased photon energy HHG will result in very large values of $\Delta k$ due to the high average degree of ionization. Quasi phase-matching under such conditions requires a very short $L_{QPM}$, which in practice is difficult to achieve, due high average ionisation. In *MMQPM* the intensity modulations are sufficiently large that Z* is significantly reduced, increasing $L_{QPM}$ and thereby allowing low-order ($m = 1$) QPM.

*MMQPM* at longer wavelengths was investigated for a vacuum coupling parameter of $\chi$~0.3, with strong enhancement observed for $q=25$ with a input laser energy of ~6mJ giving $I_{max}$~9×10$^{14}$Wcm$^{-2}$ and $p$~54mbar Ar. The HHG spectrum obtained under these conditions is shown in Figure 3a. The peak conversion efficiency per harmonic was similar to the water-window case at $>10^{-6}$ at ~32nm (25$^{th}$ order).

Allowing for filter transmission the full width half maximum (FWHM) of the quasi-phasematched harmonic comb can be estimated as ~20 harmonic orders for the water-window harmonics and 2-3 orders at the 25$^{th}$ harmonic. This narrow band enhancement is characteristic of successful phasematching, since the phase mismatch $\Delta k$ depends inversely on the harmonic order $q$. As a result, the fractional bandwidth of the enhanced harmonic spectrum ($\Delta\omega_{FWHM}$) is very similar for the two cases studied here ($\Delta\omega_{FWHM}/\omega_{190} \sim \Delta\omega_{FWHM}/\omega_{25} \sim 0.1$, where $\omega_q$ is the angular frequency of the $q^{th}$ harmonic order). Intensity tuning of the peak harmonic was observed when $I_{max}$ was varied (and thus $\Delta k$) for otherwise constant conditions. In Figure 3 the brightest harmonic is shifted from the 25$^{th}$ to the 29$^{th}$ harmonic by varying the incident energy from ~6mJ ($I_{max}$ ~$9\times10^{14}$Wcm$^2$, red dashed trace) to ~5mJ ($I_{max}$~$7\times10^{14}$Wcm$^2$, blue trace). Reducing the laser intensity causes lower ionisation and hence requires a higher harmonic order to satisfy the QPM matching condition.

Figure 4 shows the calculated enhancement in the harmonic signal as a function of Ar pressure for q = 201 and q = 25 for the conditions of Figs 2 and 3, respectively. Also shown are the measured harmonic signals normalized to the peak of the calculated curve. The calculated pressure dependence is in good agreement with the experimental data which lends support to our interpretation that the observed high efficiency of HHG arises from MMQPM. The small discrepancy between experiment and theory at high pressures is due likely to the fact that the presence of gas in the capillary causes the repartition of energy into higher modes [12] resulting in increased high frequency modulation over longer interaction lengths. This will result in a better *MMPQM* profile and a correspondingly higher signal even for poorer matching conditions.

Importantly, the fact that the pressure tuning curves are not narrow spikes at an optimum pressure for peak enhancement indicates that even random intensity profiles can see enhancement i.e. an optimal *MMQPM* intensity profile for a given value of $\Delta k$ will essentially be random for $\Delta k+\varepsilon$, where $\varepsilon$ is a small change in the value of $\Delta k$.

In conclusion, substantial enhancement of harmonics which we interpret as being due to multimode quasi-phasematching - *MMQPM* - has been observed over a narrow range of harmonic orders in argon. This approach has yielded the highest conversion efficiencies in the water-window to date with an enhancement over previous results of several orders of magnitude ($>10^3$). In principle this approach should allow the selective enhancement of harmonics across the full spectral range. Modeling suggests that substantial improvements (>10) on the current results can be achieved by optimising the coupling into the waveguide so as to predominantly excite only two modes.

MZ is grateful to the Royal Society for support through a Wolfson Merit Award, SMH to the Engineering and Physical Sciences Research Council for research funding (EP/C005449), and ML to the Rhodes Trust for a scholarship.

**References**

[1] X. F. Li, A. L'Huillier, M. Ferray, L. A. Lompr´e and G. Mainfray, *Phys. Rev. A* **39**, 5751 (1989).
[2] P. B. Corkum, *Phys. Rev. Lett.,* **71,** 1994 (1993).
[3] C. Durfee, A. Rundquist, S. Backus, C. Herne, M. M. Murnane, and H. C. Kapteyn, *Phys. Rev. Lett.,* **83**, 2187-2190 (1999).
[4] G. Tempa *et al*., Phys. Rev. Lett. **84**, 4329 (2000)
[5] E. A. Gibson *et al.*, *Phys. Rev. Lett,* **92**, 033001 (2004).
[6] M. Hentschel *et al., Nature* **414** 509 (2001)
[7] A. J. Langley *et al*., *CLF Annual Report 1999-2000*, p196, (2000)
[8] A. Paul *et al*., Nature, **421**, 51 (2003)
[9] B. Dromey *et al*. submitted Optics Express Feb 2007
[10] E. A. Gibson, *et al*., *Science,* **302**, 95–98, (2003).
[11] S. Voronov *et al*., *Phys. Rev. Lett.* , **87**, 133902 (2001).
[12] C. Courtois, A. Couairon, B. Cros, J. R. Marques G. Matthieussent, *Phys. Plas.* **8**, 3445, (2001).
[13] R. L. Abrams, *IEEE J. Quantum Electron.* **8**, 838 (1972).

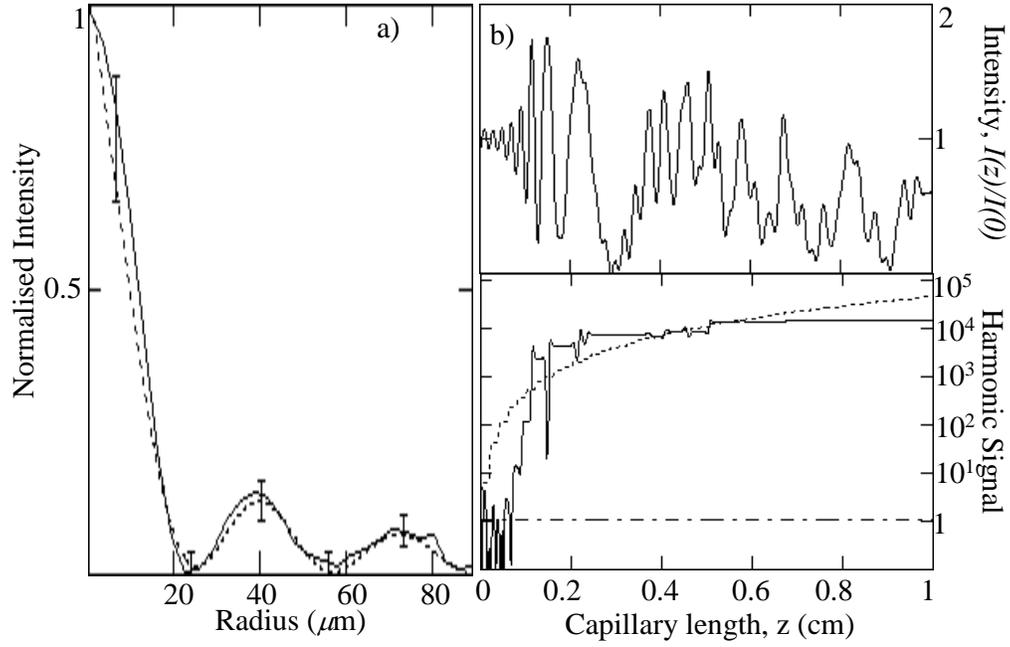

**Figure 1** The experimentally measured incident pulse profile in Figure 1a) (solid line) can be approximated by the Airy profile (dotted line) corresponding to vacuum coupling parameter of $\chi=0.2$ for a capillary with bore radius $a\sim90\mu$m (the error bars on the experimental data are due to radial averaging). The expected on-axis intensity profile, normalised at $z=0$, for a 1cm long evacuated capillary (no gas present) assuming the dotted profile in Figure 1a) is shown in Figure 1b). The resulting harmonic signal growth for $q=201$ under optimal *MMQPM* conditions ($p\sim20$mbar Ar, $I_{max}\sim1\times10^{15}$Wcm$^{-2}$, solid line) is shown in Figure 1c) normalised to that expected for a single coherence length (dashed line). Also shown (dotted line) is the expected HHG signal growth for idealised two mode beating (from Eq. 4) under optimised *MMQPM* conditions and $j=20$ ($L_{QPM} = 200\mu$m).

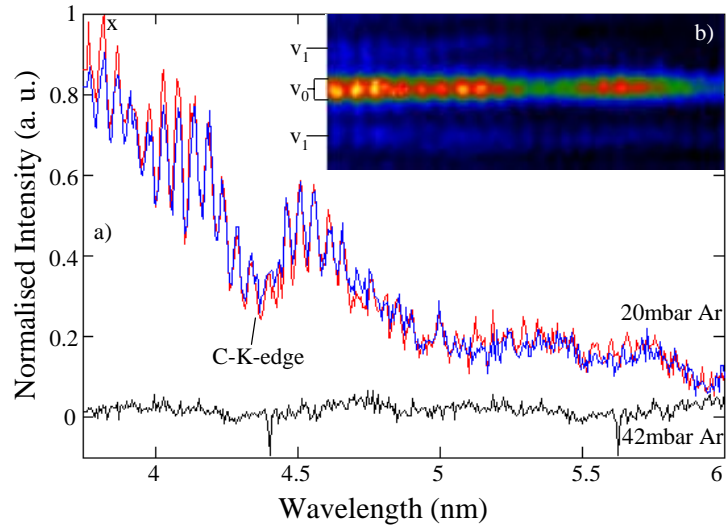

**Figure 2.** Quasi-phasematched HHG in the water-window. The red and blue trace are two separate shots recorded at $p\sim20$ mbar Ar, the black trace is for mismatched conditions with $p\sim42$ mbar Ar. Both traces (red and blue) are normalised to the value of the peak of the red trace, showing the good reproducibility of the data. The recorded spectrum is shown in the inset. A lower limit of the conversion efficiency per harmonic order is calculated by vertical integration of the background subtracted signal in the bright region over the interval '$v_0$' only. Significant signal with clear harmonic structure is present for one side lobe (marked v1 in the inset) either side of the bright central order, $v_0$. Up to six side lobes were observed experimentally and are likely due to off axis QPM.

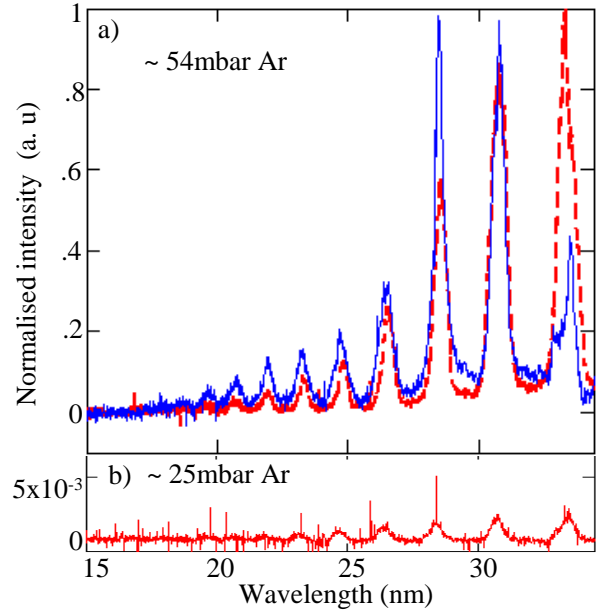

**Figure 3**. Multimode quasi-phasematching at ~30nm wavelength. a) Under matched conditions an enhancement of >200 is observed for 2-3 harmonic orders. Varying $\Delta k$ by a small change in intensity allows the position of the brightest peak to be tuned from the 25$^{th}$ to the 29$^{th}$ harmonic. $I_{max}=9\times10^{14}$ Wcm$^{-2}$ (red) and $7\times10^{14}$ (blue). Mismatched conditions can be seen in b) which shows the typical 'plateau' structure of almost equal intensities typical of HHG in the absence of phasematching [1].

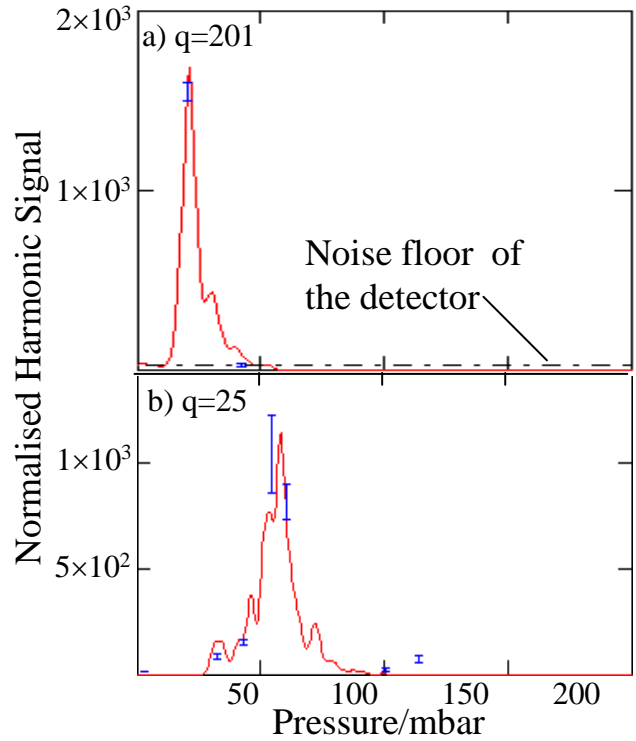

**Figure 4** Pressure tuning curves (solid lines) calculated for a) $I_{max}$~$1\times10^{15}$Wcm$^{-2}$ ($q$=201, $\chi$~0.2) and b) $I_{max}$~$9\times10^{14}$Wcm$^{-2}$ ($q$=25, $\chi$~0.3), showing expected harmonic signal enhancements over the signal expected for a single coherence length. The experimental data (points) is normalised to the peak of the pressure tuning curves. The vertical extent of the data points indicates the experimentally observed shot to shot variation in the intensity of the stated harmonic order.